 \documentclass[final,5p,times,twocolumn]{elsarticle} 

\usepackage{amssymb}
\usepackage{lipsum}
\usepackage{amsmath}
\usepackage{bm}
\usepackage{hyperref}

\usepackage{color}
\newcommand{\red}[1]{\textcolor{red}{{#1}}}

\biboptions{sort&compress}

\journal{Physics Letters B} 

\begin{document}

\begin{frontmatter}

\title{Basis Representation for Nuclear Densities from Principal Component Analysis}

\author[FZU]{Chen-Jun Lv}

\author[BUAA]{Tian-Yu Wu}

\author[FZU]{Xin-Hui Wu\corref{cor1}}

\ead{wuxinhui@fzu.edu.cn}

\author[Mila,INFN]{Gianluca Col\`o}

\author[KYT,KYTI,RIKEN]{Kouichi Hagino}


\cortext[cor1]{Corresponding Author}
\address[FZU]{Department of Physics, Fuzhou University, Fuzhou 350108, Fujian, China}
\address[BUAA]{School of Physics, Beihang University, Beijing 100191, China}
\address[Mila]{Dipartimento di Fisica, Universit\`a
degli Studi di Milano, via Celoria 16, 20133 Milano, Italy,}
\address[INFN]{INFN, Sezione di Milano, via Celoria 16, 20133 Milano, Italy}
\address[KYT]{Department of Physics, Kyoto University, Kyoto 606-8502,  Japan}
\address[KYTI]{Institute for Liberal Arts and Sciences, Kyoto University, Kyoto 606-8501, Japan}
\address[RIKEN]{RIKEN Nishina Center for Accelerator-based Science, RIKEN, Wako 351-0198, Japan}

\begin{abstract}
We develop an efficient method to represent nuclear densities using basis functions extracted via Principal Component Analysis (PCA).
Applying PCA to densities of 75 nuclei calculated with the relativistic continuum Hartree–Bogoliubov (RCHB) theory yields an orthogonal set of components that efficiently capture the dominant features of nuclear density distributions, which can be used as basis functions for nuclear density representation.
The first five basis functions account for more than 99.999\% of the total variance, demonstrating the efficiency of these PCA basis functions. 
The PCA basis achieves significantly higher accuracy and faster convergence than the Fourier-Bessel and Sum-of-Gaussians methods for reconstructing both theoretical and experimental densities. 
This approach provides an efficient and robust representation of nuclear densities, offering a practical tool for experimental density representation and for theories where densities play a central role, such as the orbital-free density functional theory, or the double folding model for nuclear reactions.

\end{abstract}


\end{frontmatter}


\section{Introduction}

Nuclear density is a fundamental quantity that represents the spatial distribution of nucleons within a nucleus, and plays an important role in nuclear many-body systems.
It determines essential ground-state properties such as charge and matter radii, surface profiles, and saturation behavior, and also governs the dynamics of nuclear reactions and collective excitations.
Actually, according to the Hohenberg-Kohn theorem~\cite{Hohenberg1964Phys.Rev.}, it encodes all nuclear many-body effects.
A flexible and convenient representation of nuclear densities is important for both experimental and theoretical studies.

On the experimental side, since the early time of Hofstadfer~\cite{Hofstadter1956Rev}, elastic electron scattering has been employed to measure the charge density distribution $\rho_{c}(r)$ of nearly all stable atomic nuclei~\cite{Vries1987ADNDT,Fricke1995ADNDT}. 
The latest experimental measurements for the unstable nucleus $^{137}$Cs can also be found in Ref.~\cite{Tsukada2023PRL}.
This technique extracts crucial information about $\rho_{c}(r)$ by analyzing the differential elastic scattering cross sections of high energy electrons off nuclei~\cite{Crannell1966Phys.Rev.,Suelzle1967Phys.Rev.,Sick1970Nucl.Phys.A,Sinha1971Phys.Lett.B,Sick1972Phys.Lett.B,Friar1973Nucl.Phys.A,Chernykh2009Phys.Rev.Lett.,Chu2009Phys.Rev.C,Adhikari2022Phys.Rev.Lett.}. 
To effectively characterize the charge density, one typically assumes that the charge density $\rho_{c}(r)$ follows specific parameterized analytical forms, and determines the corresponding parameters by fitting experimental cross-section data~\cite{Crannell1966Phys.Rev.,Suelzle1967Phys.Rev.,Sick1970Nucl.Phys.A}.
Commonly used parameterized density representations can be categorized into model-dependent and model-independent types.
Although model-dependent forms such as the modified harmonic-oscillator function (MHO)~\cite{Crannell1966Phys.Rev.,Suelzle1967Phys.Rev.,Sick1970Nucl.Phys.A,Heisenberg1970Nucl.Phys.A,Kline1973Nucl.Phys.A}, two-parameter Fermi function (2pF)~\cite{Kline1973Nucl.Phys.A,Knight1981JPGNPP,Friedrich1982Nucl.Phys.A}, three-parameter Fermi function (3pF)~\cite{Sick1970Nucl.Phys.A,Li1974Phys.Rev.C,Chu2009Phys.Rev.C}, and three-parameter Gaussian function (3pG)~\cite{Ficenec1972Phys.Lett.B,Li1974Phys.Rev.C,Ray1978Phys.Rev.C} can fit the charge density distribution reasonably well, their accuracy remains constrained by the choice of predefined form~\cite{Dreher1974Nucl.Phys.A,Wu2025ADNDT}. 
In contrast, model-independent analysis like Fourier-Bessel (FB)~\cite{Dreher1974Nucl.Phys.A,Starodubsky1994Phys.Rev.C} and Sum-of-Gaussians (SOG) methods~\cite{Sick1974Nucl.Phys.A,Cavedon1982Phys.Lett.B,Liu2020Phys.Lett.B} significantly improve the fitting accuracy of charge density distributions~\cite{Vries1987ADNDT}.
However, as a parameter-fitting method, the SOG method has a complex fitting procedure that requires multiple iterations.
Its performance heavily relies on the selection and simultaneous optimization of numerous parameters, which can not always consistently converge to an optimal solution.
The FB method, as a basis expansion approach, facilitates obtaining expansion coefficients from experimental data, but the determination of cutoff radius primarily relies on empirical judgment and lacks a unified standard. 
The determination of FB coefficients typically requires measurement at high momentum transfer, which often necessitates extrapolation from low momentum transfer data, thereby introducing additional uncertainties~\cite{Wu2025ADNDT}.

On the theoretical side, in conventional quantum mechanical approaches, the wave function serves as the primary degree of freedom to represent quantum many-body systems, and a wide range of systematic basis-representation techniques have been developed to describe it accurately.
In contrast, density functional theory replaces the many-body wave function with the one-body density as the fundamental variable.
Despite this conceptual shift, the development of systematic and well-controlled basis representations for densities has received comparatively little attention.
Existing studies typically represent densities on discrete spatial grids or reconstruct them indirectly through summations over occupied single-particle orbitals.
Grid-based discretization of densities is employed for orbital-free density functional theory (DFT)~\cite{Wu2022Phys.Rev.C, Colo2023PTEP, Hizawa2023PRC, Chen2024IJMPE, Wu2025CP, Wu2025PRC, Wu2025arXiv}, which can capture detailed spatial features but incurs additional computational overhead compared to basis-set expansion methods and provides limited analytical insight.
On the other hand, orbital summation, though widely used in mean-field or Kohn-Sham DFT~\cite{Bender2003Rev.Mod.Phys., Meng2016book, Colo2020nuclear}, effectively reverts to a wave-function-based description rather than treating the density as an independent fundamental variable.
For very large systems, a wave-function-based description is computationally impractical.
Consequently, a systematic representation of the density is helpful for the further development of DFT, which is particularly valuable for orbital-free implementations, where the density itself serves as the sole basic variable~\cite{Levy1984PRA}.

In this work, we propose a systematic framework for constructing basis representations of nuclear densities using the Principal Component Analysis (PCA) method~\cite{Jolliffe2002principal}.
The goal is to identify a compact set of basis functions that efficiently capture both the global and local features of the density.
Within this framework, the nuclear density is effectively represented as a linear combination of orthogonal principal components, which are derived from covariance patterns among a representative ensemble of density distributions.

\section{Density bases from principal component analysis}\label{sec:theo}

Principal component analysis is a technique designed to extract a set of orthogonal principal components that capture the maximum variances within a dataset~\cite{Jolliffe2002principal}.
This method employs a transformation to convert the original variables into a new set of variables known as principal components (PCs).
These components are the eigenvectors of the covariance matrix: as such, they are mutually orthogonal and they are ordered according to the magnitude of the corresponding eigenvalues, which quantify the importance of each component. 
Consequently, only the first few PCs are needed to capture the majority of significant features presented in the original data, enabling effective dimensionality reduction while maintaining the integrity of essential information.
This can in principle yield the optimal basis functions for representing the original variables.

Here we employ PCA to identify a set of principal density patterns from a collection of distinct nuclear density distributions.
These mutually orthogonal principal density patterns are then used to construct a concise set of basis functions, enabling an efficient representation of various nuclear density distributions. 
We begin with a collection of $m$ distinct nuclear density distributions of different nuclei under the spherical assumption.
The $k$-th nuclear density distribution is denoted as $\rho_{k}(r_{i})$, where $i=1,...,n$ indicates that the radial coordinate is discretized into $n$ equidistant grid points.
Accordingly, each density distribution can also be represented as an $n$-dimensional vector $\boldsymbol{\rho}_{k}$.
These density vectors are stacked to form the matrix,
\begin{align}
    \boldsymbol{X} =\begin{bmatrix}
    \boldsymbol{\rho}_{1} \\
    \boldsymbol{\rho}_{2} \\
    ...\\
    \boldsymbol{\rho}_{m} \\
   \end{bmatrix} 
       = \begin{bmatrix}
    \rho_{1}(r_{1}) & \rho_{1}(r_{2}) & ... & \rho_{1}(r_{n})\\
    \rho_{2}(r_{1}) & \rho_{2}(r_{2}) & ... & \rho_{2}(r_{n})\\
    ...    & ...&...&...\\
    \rho_{m}(r_{1}) & \rho_{m}(r_{2}) & ... & \rho_{m}(r_{n})
  \end{bmatrix} .
  \label{eq:1}
\end{align}
In this matrix, the matrix element $\rho_k(r_i)$ represents the density value of the $k$-th nucleus at the $i$-th radial grid point.
Thus, each row contains the density values of a specific density distribution across all $n$ radial grid points, while each column contains the density values of all $m$ density distributions at the same radial grid point.
The covariance matrix is then constructed as follows,
\begin{align}
\boldsymbol{C}=\frac{1}{m-1}\boldsymbol{X}^{T}\boldsymbol{X} .
\label{eq:2}
\end{align}
This covariance matrix $\boldsymbol{C}$ captures the joint variation of density values at different radial coordinates, and its eigenvectors define an orthogonal basis in the space of density distributions.
We achieve the diagonalization of the covariance matrix by solving its eigenvalue equation 
\begin{align}
\boldsymbol{C}v_{j}=\lambda_{j}v_{j}.\label{eq:3}
\end{align}
The obtained eigenvectors $v_{j}$ form the extracted orthogonal PCs, while the eigenvalues represent the importance of the corresponding PC.

In a recent study, one of the authors derived the radial point proton density distributions $\rho_{p}(r)$ of 130 stable nuclei by expanding the charge density $\rho_{c}(r)$ measured through elastic electron scattering experiments~\cite{Wu2025ADNDT}.
Among these, the proton density distributions of 75 nuclei were expanded using the model-independent method.
These distributions are regarded as high-precision proton densities from experiments.
To ensure comparability with experimental data, we have calculated the proton densities of these nuclei using the relativistic continuum Hartree-Bogoliubov theory (RCHB)~\cite{Meng1998NPA, Xia2018ADNDT} with the PC-PK1 functional~\cite{Zhao2010Phys.Rev.C}, and established them as the benchmark dataset for PCA.
In the calculation, spherical assumption is adopted to solve the RCHB equations, even though some of the nuclei are deformed in the ground state. Notice that our primary purpose is to generate typical density distributions for a given mass region, rather than precisely reproduce the experimental data.
A list of these 75 nuclei can be seen in the Tab.~S1 of Supplemental Material~\cite{supplement}.
The proton density distribution function $\rho_{p}(r)$ is defined over 101 radial grid points covering radial range of $r=0\sim10$ fm, resulting in a raw matrix $\boldsymbol{X}$ of dimensions $75\times 101$. 
Following the method described above, the $j$-th PC $v_{j}$ can be computed.
One can thus obtain 75 orthogonal principal density distribution vectors $\boldsymbol{v}_{1}, \boldsymbol{v}_{2}, …, \boldsymbol{v}_{75}$.
The importance of each PC is determined by its eigenvalue $\lambda_{j}$, which is listed in descending order of magnitude.
Thus, $\boldsymbol{v}_{1}$ represents the most significant feature of the original vectors, followed by $\boldsymbol{v}_{2}$, and $\boldsymbol{v}_{75}$ the least important.
Due to their mutual orthonormality, PCs can efficiently analyze the commonalities and differences in nuclear density distributions and serve as the basis functions for density representation.
The density distribution of an atomic nucleus can be described through their linear combination, 
\begin{align}
\boldsymbol{\rho}\approx\sum_{j=1}^{t}a_{j}\boldsymbol{v}_{j} , 
\label{eq:4}
\end{align}
where $t$ denotes the selected number of PCs. 
$a_{j}$ represents the coefficient corresponding to each PC basis vector, which is the inner product of the nuclear density vector and the PC vector, $a_{j}=\langle \boldsymbol{\rho},\boldsymbol{v}_{j} \rangle$.

Note that to eliminate the scale effects arising from variations in mass number $A$ and proton number $Z$ across different nuclei, we normalized density distributions before PCA.
This allows the PCA to focus on extracting the intrinsic shape characteristics.
To this end, using $^{40}\text{Ca}$ as the standard, we applied scaling factors simultaneously to the radial coordinate and the density amplitude. 
This normalizes the root-mean-square radius of the proton density distribution for each nucleus to 4.104 fm and constrains the total proton number to 20.
The detailed procedure of scaling and its reversal is provided in the Supplementary Material~\cite{supplement}.
This procedure ensures that the established set of density basis functions is applicable to all atomic nuclei.

\section{Results and discussions}

Following the procedure in Section~\ref{sec:theo}, we obtained 75 orthogonal principal components of density distribution patterns, which are sorted by importance based on the magnitude of their eigenvalues.
They serve as a set of basis functions for density representation.
Using all of them as a basis to describe the nuclear density distribution would be cumbersome, and one should thus select only the first several ones.
It is therefore necessary to determine a proper number $t$ of PCs for characterizing the nuclear density distribution (see Eq.~\eqref{eq:4}). 

\begin{figure}[!htbp]
  \centering
  \includegraphics[width=0.95\linewidth]{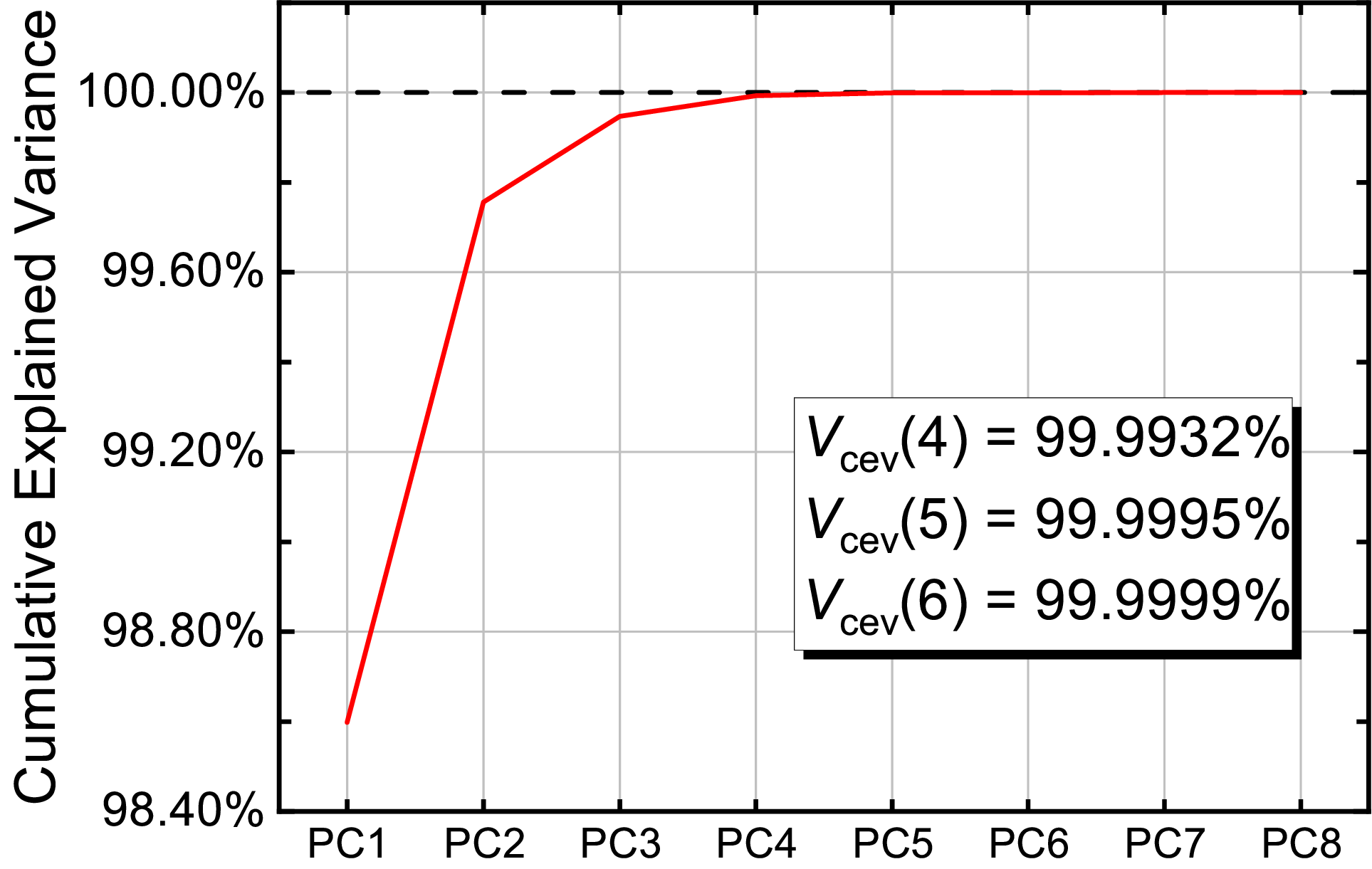}
  \caption{Cumulative explained variance ratio curves for PCs derived from proton density distributions of 75 nuclei with PCA.
  The specific values up to PC4, PC5, and PC6 are indicated in the figure.
  }
  \label{Fig:1}
\end{figure}

To this end, we have calculated the explained variance ($V_{\rm ev}$) of each PC.
This metric, defined as the ratio of the variance captured by a single PC to the total variance, measures each component's contribution to the total information in the dataset. 
Since the variance is the normalized eigenvalue of the covariance matrix, the explained variance is calculated as follows:
\begin{align}
 V_{\rm ev}(j) =\frac{\lambda_{j}}{\sum_{i=1}^{m}\lambda_{i}}.
\label{eq:5}
\end{align}
Here, $m=75$ is the total number of PCs.
The cumulative explained variance is then defined as
\begin{align}
 V_{\rm cev}(t) = \sum_{j=1}^{t} V_{\rm ev}(j) = \frac{\sum_{j=1}^{t}\lambda_{j}}{\sum_{i=1}^{m}\lambda_{i}},
\label{eq:6}
\end{align}
which represents the sum of the first $t$ density basis functions' contribution to the total information in the density representation.
The cumulative explained variance ratio curve is shown in Fig.~\ref{Fig:1}.

\begin{figure}[!htbp]
  \centering
  \includegraphics[width=0.95\linewidth]{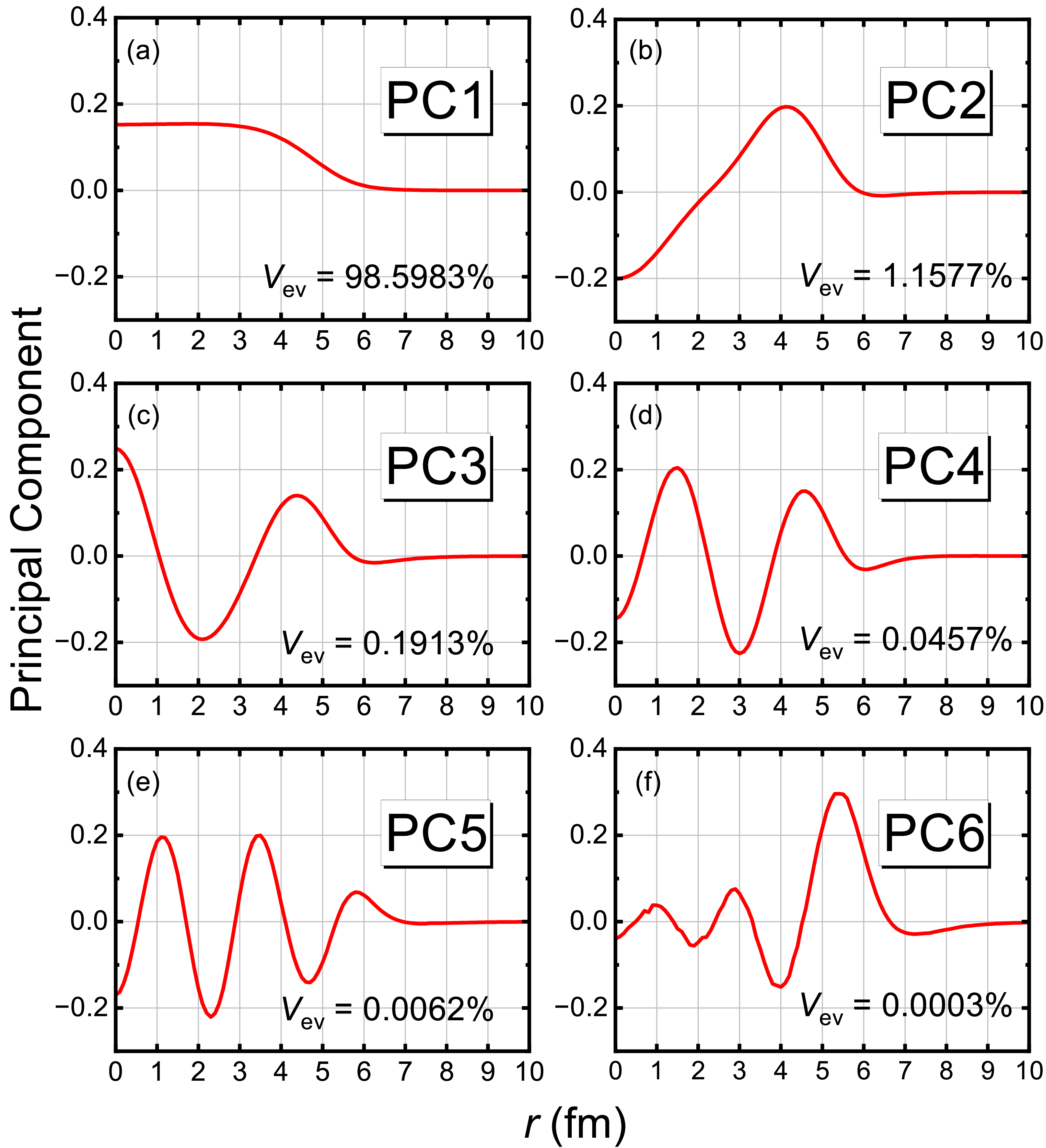}
  \caption{Radial spatial distributions of principal components PC1-PC6 for the nuclear proton density distribution, with explained variance ratio indicated for each component.
  }
  \label{Fig:2}
\end{figure}

One can see that the explained variance of PC1 is already as large as 98.60\%.
This is reasonable as the shapes of the nucleon density distributions for different nuclei are roughly similar after scaling due to the saturation of nuclear forces, with the exception of very light nuclei.
The densities are roughly around the saturation value in the interior region, and decay exponentially in the surface region.
Therefore, a density distribution function in the form of a Fermi distribution can well describe the main characteristics of nuclear density [see Fig.~\ref{Fig:2} (a)].
It can be observed that the cumulative explained variance ratio of the first 5 PCs has reached as high as 99.9995\%, which means that the first 5 basis functions from PCA are already sufficient to characterize the dominant information of proton density distributions of atomic nuclei.

The first six radial proton density basis functions from PCA are presented in Fig.~\ref{Fig:2}.
As mentioned before, the PC1 [Fig.~\ref{Fig:2}(a)] represents the dominant similar features of all densities, i.e., a flat center and a tail that asymptotically approaches zero, which is usually represented in the form of a Fermi distribution.
The large explained variance ratio of this PC1 further confirms the widespread presence of such universal characteristics across different nuclear density distributions.
It includes the main characteristics of the density distribution, except those related to the quantum shell effects.
Higher-order principal components (PC2-PC6) demonstrate markedly different characteristics, i.e., they possess smaller explained variance ratios and exhibit oscillatory structures with multiple nodes as in Figs.~\ref{Fig:2}~(b)-(f).
They represent detailed features of the density that might arise from shell effects, surface diffuseness effects and others.    
For example, PC2 exhibits a pronounced central dip, potentially reflecting the central depression phenomenon (i.e., bubble structure) observed in density distributions of some nuclei.
Different oscillatory structures of other PCs could reflect the oscillations in density distributions, which arise from the different major shells of atomic nuclei.

This set of principal components from PCA can serve as basis vectors for representing nuclear proton densities as shown in Eq.~\eqref{eq:4}.
The inclusion of more principal components leads to a more accurate representation of the densities, see the Fig.~S1 of Supplementary Material~\cite{supplement} for details.
To evaluate the performance and characteristics of this proposed procedure for representing nuclear proton densities, we compare it with two widely-employed model-independent approaches, i.e., the Fourier-Bessel (FB) expansion method and the Sum-of-Gaussians (SOG) fitting method. 
The FB expansion expands proton density as a linear combination of a complete set of basis functions, i.e., the Bessel function of order zero $j_{0}(x)=\frac{\sin x}{x}$ ~\cite{Dreher1974Nucl.Phys.A},
\begin{align}
\rho(r) = 
\begin{cases}
    \sum_{v=1}^{N_{\rm FB}}a_{v}j_{0}(\frac{v\pi r}{R_{cut}}) & \text{for } r\leq R_{\rm cut}, \\
    0 & \text{for } r> R_{\rm cut},
\end{cases}
\label{eq:7}
\end{align}
where $R_{\rm cut}$ is the cutoff radius of the density distribution.
$N_{\rm FB}$ is the number of basis functions that are adopted, and $a_v$ is the expansion coefficient.
In the SOG fitting method, the nuclear density distribution $\rho(r)$ is expressed as a superposition of multiple Gaussian functions~\cite{Sick1974Nucl.Phys.A},
\begin{align}
\rho(r)=\frac{Z}{2\pi^{3/2}\gamma^{3}}\sum_{i=1}^{N_{g}}\frac{Q_{i}}{1+2R^{2}_{i}/\gamma^{2}}\times\left[e^{-(\frac{r-R_{i}}{\gamma})^{2}}+e^{-(\frac{r+R_{i}}{\gamma})^{2}}\right],
\label{eq:8}
\end{align}
in which $N_{g}$ represents the total number of Gaussians, $\gamma$ denotes the common width of the Gaussians, $R_{i}\in [0,R_{\rm max}]$ are positions to be determined, and $Q_i$ are amplitudes.

The PCA method characterizes nuclear density through linear superposition of basis functions (PCs) extracted from extensive nuclear proton density distributions. 
The SOG method employs a summation of Gaussian functions, and the FB approach utilizes a linear combination of Bessel functions.
They all represent nuclear density by combining several terms.
However, different numbers of fitting parameters are included in each term among these three different methods.
According to Eq.\eqref{eq:4}, $N$ fitting terms in the PCA method correspond directly to $N$ parameters $a_{j}$ to be fitted. 
As seen in Eq.\eqref{eq:7}, the FB expansion method requires fitting $N$ Bessel function coefficients $a_{v}$ plus the cutoff radius $R_{\rm cut}$, resulting in a total of $N+1$ fitting parameters.
For the SOG method, as shown in Eq.\eqref{eq:8}, each Gaussian function involves two parameters, the Gaussian center position $R_{i}$ and its corresponding amplitude $Q_{i}$, in addition to a shared Gaussian width parameter $\gamma$.
Thus, $N$ fitting terms correspond to $2N+1$ fitting parameters. 
We summarize the number of parameters corresponding to the number of fitting terms in Tab.~\ref{tab:1} for these three methods.
The number of fitting parameters for the SOG method is typically double than that of the PCA and FB methods for the same included terms.
It should be noted that to maintain consistency with the PCA method, all the FB and SOG parameters are likewise obtained by fitting the density distribution in the coordinate space, which slightly differs from parameters determined in experimental analyses~\cite{Sick1974Nucl.Phys.A, Dreher1974Nucl.Phys.A}.

Furthermore, both PCA and FB methods belong to the category of basis function expansion methods.
The corresponding expansion coefficients are obtained by computing inner products between the target density function and a predefined set of basis functions. 
This process ensures systematic procedures to obtain the optimal parameters. 
In contrast, the SOG method is a fitting-based method.
It requires determining numerous parameters through an optimization process, including the positions $R_{i}$, amplitudes $Q_{i}$ of the Gaussian functions, and a unified width parameter $\gamma$. 
This process is inherently unstable, as one may not always find the optimal parameters in the fitting procedure which includes numerous parameters.

\begin{table}[h]
	\centering
    \resizebox{0.45\textwidth}{!}{
	\begin{tabular}{@{}cccccccccc@{}}
		\hline
		\multicolumn{2}{c}{Number of terms}     & 1 & 2 & 3 & 4 & 5  & 6  & 7  & 8  \\ \hline
		                          & PCA           & 1 & 2 & 3 & 4 & 5  & 6  & 7  & 8  \\
		Number of Parameters    & SOG           & 3 & 5 & 7 & 9 & 11 & 13 & 15 & 17 \\
		                        & FB            & 2 & 3 & 4 & 5 & 6  & 7  & 8  & 9  \\ \hline
	\end{tabular}
    }
	\caption{The numbers of parameters corresponding to numbers of fitting terms for the PCA, SOG, and FB methods.}
	\label{tab:1}
\end{table}

\begin{figure}[!t]
  \centering
  \includegraphics[width=0.95\linewidth]{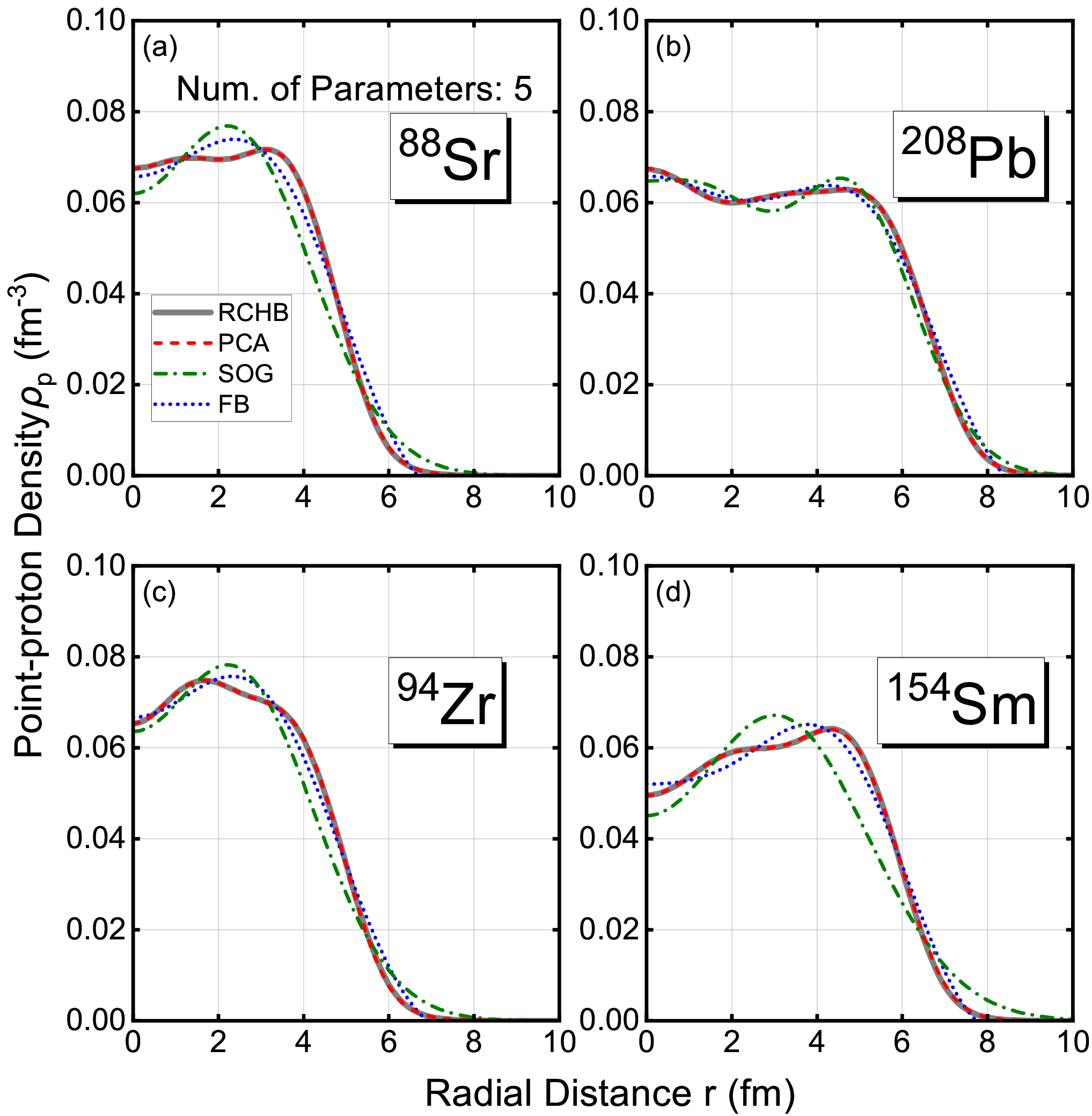}
  \caption{Comparison of the representation performances for proton density distributions of $^{88}$Sr, $^{208}$Pb, $^{94}$Zr, and $^{154}$Sm from RCHB calculations by the PCA, SOG, and FB methods with the number of fitting parameters fixed at 5.
  }
  \label{Fig.3}
\end{figure}

Figure~\ref{Fig.3} presents a comparison of the reconstructed nuclear density distributions obtained by the PCA, SOG, and FB methods for four nuclei under the condition of using the same 5 fitting parameters in these three methods.
The target densities of $^{88}$Sr, $^{208}$Pb, $^{94}$Zr, and $^{154}$Sm nuclei are obtained from the RCHB calculations.
Note that the densities of all four nuclei are included in the set of 75 densities used in the PCA procedure.

As can be seen in Fig.~\ref{Fig.3}, in the case of using the same numbers of parameters, i.e., 5 parameters, the PCA method works significantly better than the other two methods in representing proton densities of all these four nuclei.
With 5 parameters, only the densities represented by the PCA method can nicely reproduce the target proton densities from the RCHB calculations, while obvious deviations can be seen for the densities from the SOG and FB methods.
This indicates the effectiveness of using PCA method to build basis functions for nuclear densities.

\begin{figure}[!h]
  \centering
  \includegraphics[width=0.95\linewidth]{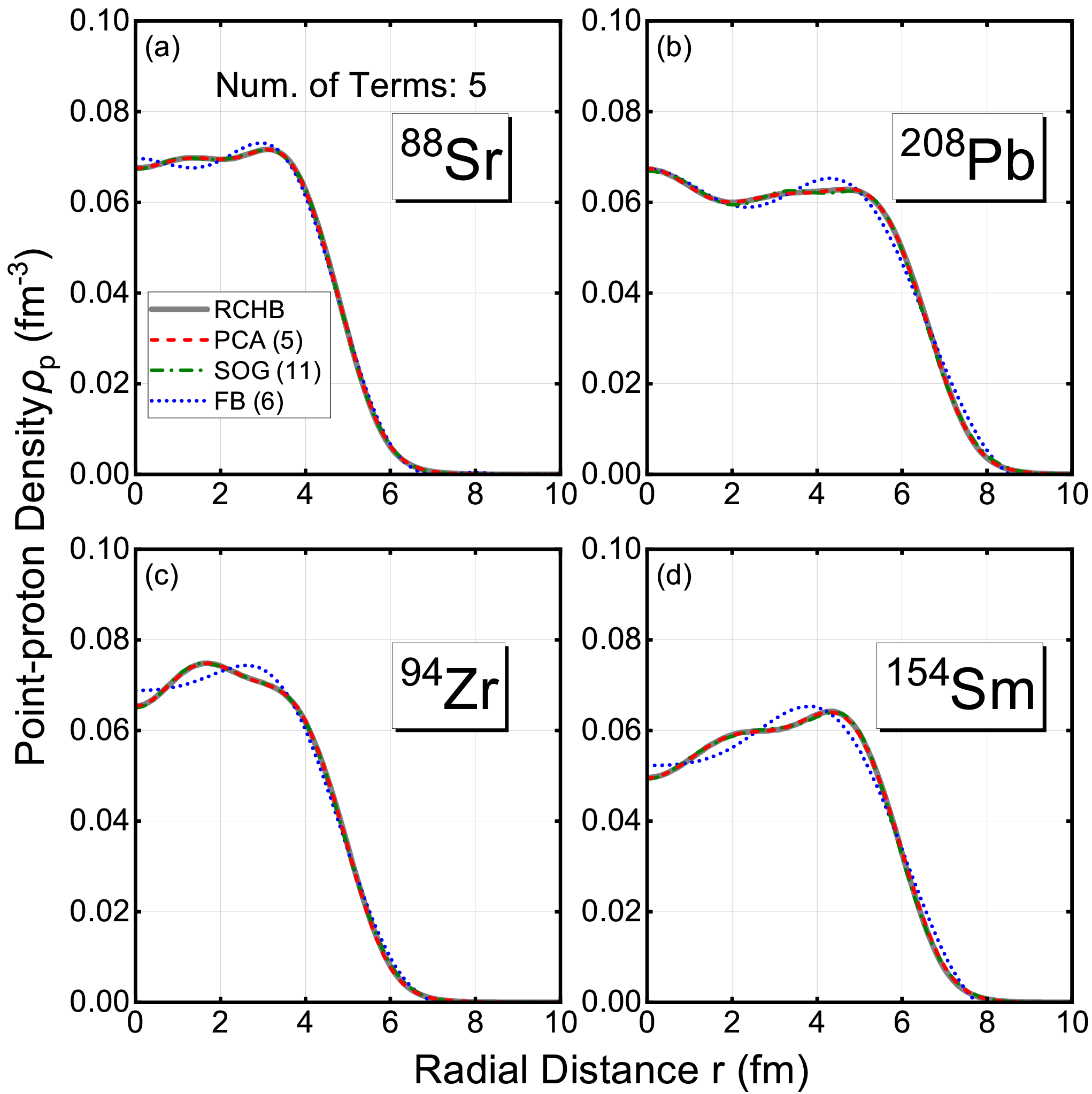}
  \caption{Comparison of the representation performances for proton density distributions of $^{88}$Sr, $^{208}$Pb, $^{94}$Zr, and $^{154}$Sm from RCHB calculations by the PCA, SOG, and FB methods with the number of fitting terms fixed at 5.
  The numbers in the bracket of the caption represent the number of fitting parameters included in each method.
  }
  \label{Fig.4}
\end{figure}

Figure~\ref{Fig.4} presents the comparison of the reconstructed nuclear density distributions obtained by the PCA, SOG, and FB methods under the condition of using the same 5 fitting terms.
In this case, one can still see obvious deviations between the FB densities and the target RCHB ones.
The SOG method can now also well reproduce the target densities of these four nuclei from the RCHB calculations.
However, it should be noted that the number of fitting parameters included in the SOG method is significantly larger than the other two methods when the numbers of fitting terms are the same.
The densities plotted on a logarithmic scale, corresponding to the densities presented in Figs.~\ref{Fig.3} and \ref{Fig.4}, are provided in the Figs.~S2 and S3 of Supplemental Material~\cite{supplement} to facilitate a detailed comparison of the density tails.

To quantitatively compare performances of these three methods in representing the target RCHB density distribution in Figs.~\ref{Fig.3} and \ref{Fig.4}, the relative integrated density difference $\Delta_{\rm Err.}$ is defined as follows,
\begin{align}
  \Delta_{\rm Err.}=\frac{\int_{0}^{\infty}{\rm d}r~4\pi r^{2}|\rho_{\rm RCHB}(r)-\rho_{\rm Fit}(r)|}{\int_{0}^{\infty}{\rm d}r~4\pi r^{2}\rho_{\rm RCHB}(r)},
\label{eq.Err}
\end{align}
where $\rho_{\rm Fit}$ denotes the densities represented by the PCA, FB, or SOG respectively.
The results are presented in Tab.~\ref{tab.Err}. 
It can be observed that the PCA method achieves higher accuracy than the other two methods in both the fixed number of parameters and fixed number of terms cases.

\begin{table}[!t]
	\centering
    \resizebox{\columnwidth}{!}{
	\begin{tabular}{ccccccc}
		\hline
		           & \multicolumn{3}{c}{Fixed Num. of Parameters}         & \multicolumn{3}{c}{Fixed Num. of Terms}         \\ \hline
		Nuclei     & PCA          & FB           & SOG          & PCA          & FB           & SOG          \\ \hline
		$^{88}$Sr  & 0.004     & 0.207     & 0.217     & 0.004     & 0.037     & 0.005     \\
		$^{208}$Pb & 0.004     & 0.130     & 0.084     & 0.004     & 0.101     & 0.015     \\
		$^{94}$Zr  & 0.009     & 0.159     & 0.190     & 0.009     & 0.067     & 0.009     \\
		$^{154}$Sm & 0.011     & 0.129     & 0.265     & 0.011     & 0.123     & 0.012     \\ \hline
	\end{tabular}
    }
	\caption{Relative integrated density difference $\Delta_{\rm Err.}$ between the RCHB densities and the densities represented by the PCA, FB, and SOG methods. The densities data are the same as in Figs.~\ref{Fig.3} and \ref{Fig.4}.}
	\label{tab.Err}
\end{table}

To comprehensively compare the performances of the PCA, SOG, and FB methods in characterizing nuclear density distributions, we perform the comparisons for the densities of all the 75 nuclei.
For the SOG method, as mentioned, the included numbers of parameters are typically double that of the FB and PCA methods with the same included terms, and thus its slow convergence speed is expected.
The FB and PCA methods are both ``basis representation'' methods.
Therefore, the comparison indicates the different effectiveness of these two sets of basis functions.
The limitation of the FB method comes from the fixed form of the Bessel function basis set, whose functional forms are independent of the intrinsic characteristics of density distributions, making it difficult to accurately describe nuclear densities with only few basis functions.
On the contrary, the PCA method employs basis functions extracted as principal components from extensive nuclear density distributions. 
These bases inherently contain useful density distribution information, enabling PCA to achieve optimal fitting performance with only a few parameters.

\begin{figure}[h]
  \centering
  \includegraphics[width=0.95\linewidth]{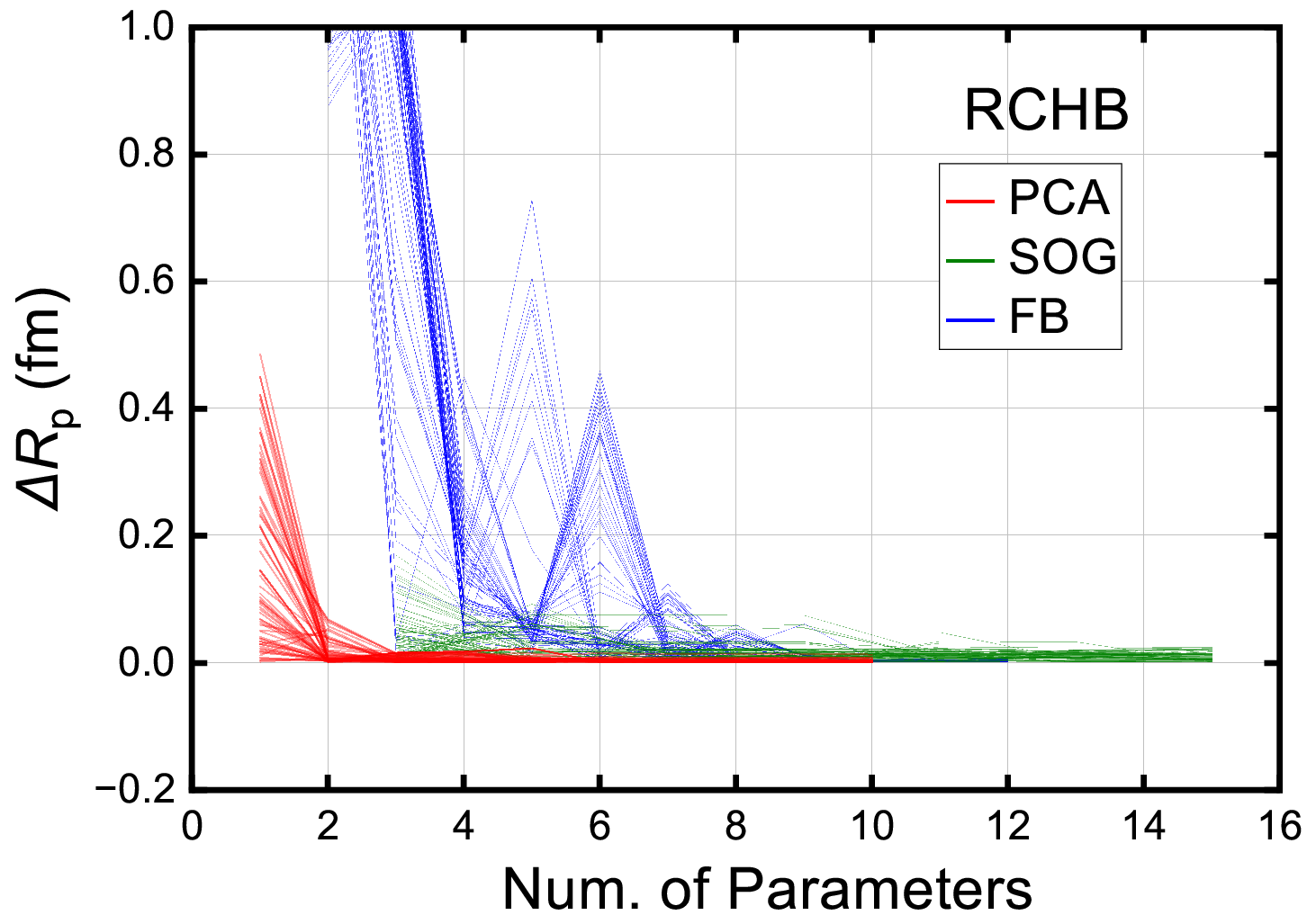}
  \caption{The absolute differences between the proton root-mean-square radii from the represented densities of the PCA, SOG, and FB methods and the target values from the RCHB calculations of 75 adopted nuclei.
  The differences under the cases with different numbers of parameters adopted in these three methods are presented.
  }
  \label{Fig.5}
\end{figure}

\begin{figure}[h]
  \centering
  \includegraphics[width=0.95\linewidth]{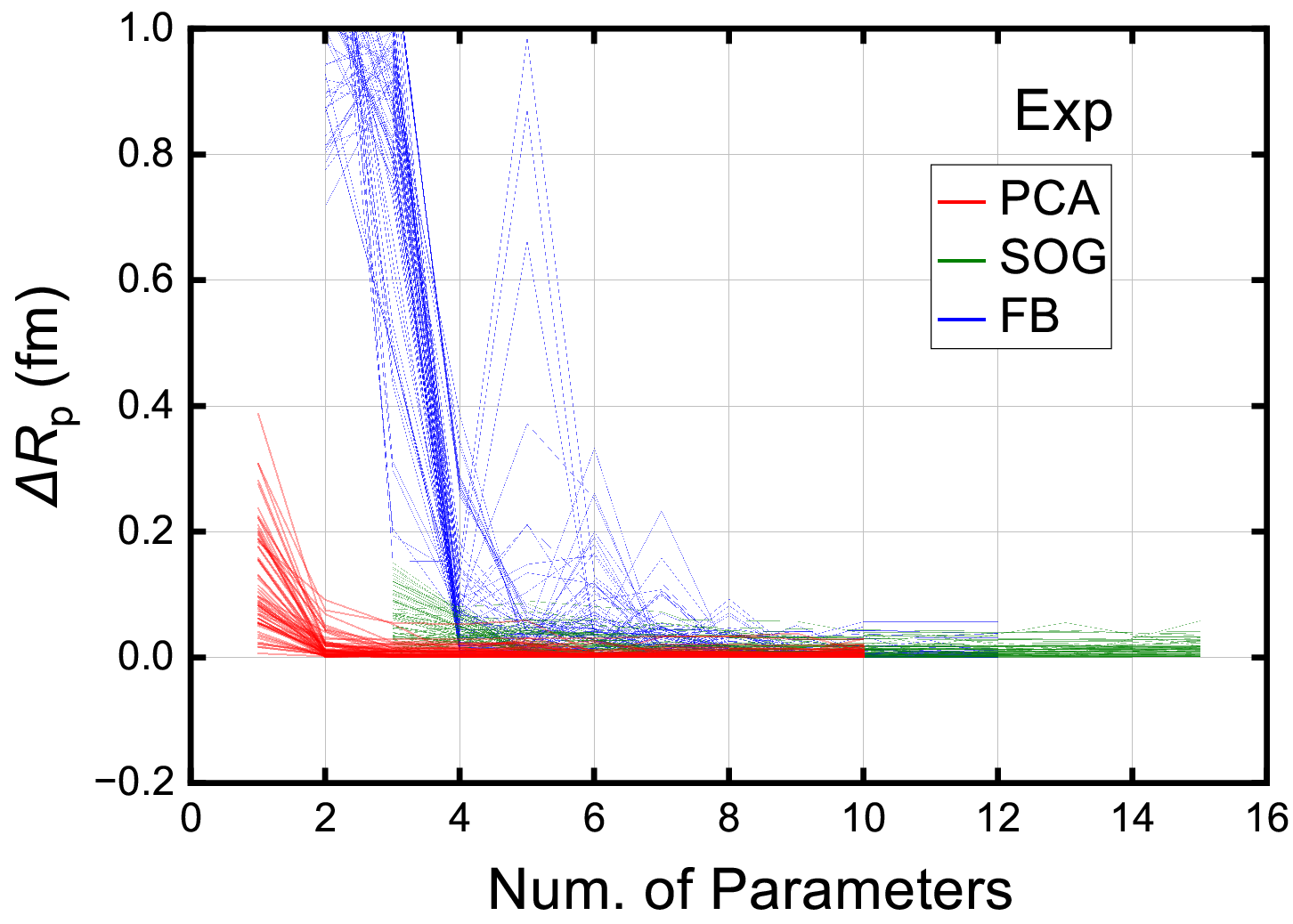}
  \caption{The absolute differences between the proton root-mean-square radii from the represented densities of the PCA, SOG, and FB methods and the target values from the experimental dates in Ref.~\cite{Wu2025ADNDT}  of 75 adopted nuclei.
  The differences under the cases with different numbers of parameters adopted in these three methods are presented.
  }
  \label{Fig.6}
\end{figure}

The performances of the PCA, SOG, and FB methods in representing proton density distributions from the experiments~\cite{Wu2025ADNDT} are also compared in Fig.~\ref{Fig.6}.
Similar conclusions with the case of representing theoretical densities are obtained.
This indicates that the density bases from principal component analysis work well in representing both theoretical and experimental densities.
Note that these PCA density bases are also proved to be universal in representing densities for nuclei across the full nuclear chart, see the Figs.~S4-S6 of Supplemental Material~\cite{supplement} for details.

\section{Summary}

This work presents a systematic framework for constructing efficient basis representations of nuclear densities using Principal Component Analysis (PCA).
By applying PCA to density distributions of 75 nuclei calculated from the relativistic continuum Hartree-Bogoliubov (RCHB) theoretical calculations, a set of orthogonal basis functions that capture the principal features of nuclear densities is obtained.
These principal components are ordered by explained variance, with the first five components accounting for over 99.999\% of the total variance.
When used to reconstruct nuclear densities, the PCA basis significantly outperforms conventional methods, namely Fourier-Bessel expansion and Sum-of-Gaussians fitting, in both accuracy and parameter efficiency, for both theoretical and experimental densities. 
With only five parameters, PCA basis reproduces reference densities with negligible deviation.
A systematic comparison of rms radii for 75 nuclei further shows that PCA basis achieves the fastest convergence among the three methods.
This approach offers an efficient and robust representation of nuclear densities, provides a useful representation for density measurement experiments, and serves as a practical tool for advancing density-centric theoretical frameworks such as orbital-free density functional theory, or reaction models that use folding of potentials with densities.

\section*{Acknowledgments}

This work was partly supported by the National Natural Science Foundation of China (Grant No. 12405134), the China Postdoctoral Science Foundation (Grant No. 2021M700256), and the start-up grant XRC-23103 of Fuzhou University.

\bibliographystyle{elsarticle-num}
\bibliography{paper}

\clearpage

\end{document}


\begin{frontmatter}

\title{Supplementary Materials for "Basis Representation for Nuclear Densities from Principal Component Analysis"}

\author[FZU]{Chen-Jun Lv}

\author[BUAA]{Tian-Yu Wu}

\author[FZU]{Xin-Hui Wu\corref{cor1}}

\ead{wuxinhui@fzu.edu.cn}

\author[Mila,INFN]{Gianluca Col\`o}

\author[KYT,KYTI,RIKEN]{Kouichi Hagino}


\cortext[cor1]{Corresponding Author}
\address[FZU]{Department of Physics, Fuzhou University, Fuzhou 350108, Fujian, China}
\address[BUAA]{School of Physics, Beihang University, Beijing 100191, China}
\address[Mila]{Dipartimento di Fisica, Universit\`a
degli Studi di Milano, via Celoria 16, 20133 Milano, Italy,}
\address[INFN]{INFN, Sezione di Milano, via Celoria 16, 20133 Milano, Italy}
\address[KYT]{Department of Physics, Kyoto University, Kyoto 606-8502,  Japan}
\address[KYTI]{Institute for Liberal Arts and Sciences, Kyoto University, Kyoto 606-8501, Japan}
\address[RIKEN]{RIKEN Nishina Center for Accelerator-based Science, RIKEN, Wako 351-0198, Japan}

\end{frontmatter}


\section*{List of 75 nuclei included in the PCA prodedure}

The list of 75 nuclei, whose density distributions are adopted in the PCA procedure in the main text, are presented in Tab.~\ref{suptable:1}.
One can see that these nuclei are widely distributed on the nuclide chart (see also Fig.~\ref{Fig.7}), basically covering nuclei at various positions along the stability line.

\begin{table}[!htp]
	\centering
	\begin{tabular}{ c c c c c }
		\hline
        \multicolumn{5}{c}{\textbf{List of 75 nuclei}} \\
		\hline
		${}^{12}$C   & ${}^{15}$N   & ${}^{16}$O   & ${}^{24}$Mg  & ${}^{26}$Mg  \\
		${}^{27}$Al  & ${}^{28}$Si  & ${}^{29}$Si  & ${}^{30}$Si  & ${}^{31}$P   \\
		${}^{32}$S   & ${}^{34}$S   & ${}^{36}$S   & ${}^{40}$Ar  & ${}^{39}$K   \\
		${}^{40}$Ca  & ${}^{48}$Ca  & ${}^{48}$Ti  & ${}^{50}$Ti  & ${}^{50}$Cr  \\
		${}^{52}$Cr  & ${}^{54}$Cr  & ${}^{54}$Fe  & ${}^{56}$Fe  & ${}^{58}$Fe  \\
		${}^{59}$Co  & ${}^{58}$Ni  & ${}^{60}$Ni  & ${}^{62}$Ni  & ${}^{64}$Ni  \\
		${}^{63}$Cu  & ${}^{65}$Cu  & ${}^{64}$Zn  & ${}^{66}$Zn  & ${}^{68}$Zn  \\
		${}^{70}$Zn  & ${}^{70}$Ge  & ${}^{72}$Ge  & ${}^{74}$Ge  & ${}^{76}$Ge  \\
		${}^{88}$Sr  & ${}^{90}$Zr  & ${}^{92}$Zr  & ${}^{94}$Zr  & ${}^{92}$Mo  \\
		${}^{94}$Mo  & ${}^{96}$Mo  & ${}^{98}$Mo  & ${}^{100}$Mo & ${}^{104}$Pd \\
		${}^{106}$Pd & ${}^{108}$Pd & ${}^{110}$Pd & ${}^{116}$Sn & ${}^{124}$Sn \\
		${}^{144}$Sm & ${}^{148}$Sm & ${}^{150}$Sm & ${}^{152}$Sm & ${}^{154}$Sm \\
		${}^{154}$Gd & ${}^{158}$Gd & ${}^{166}$Er & ${}^{174}$Yb & ${}^{175}$Lu \\
		${}^{192}$Os & ${}^{196}$Pt & ${}^{204}$Hg & ${}^{203}$Tl & ${}^{205}$Tl \\
		${}^{204}$Pb & ${}^{206}$Pb & ${}^{207}$Pb & ${}^{208}$Pb & ${}^{209}$Bi \\
		\hline
	\end{tabular}
	\caption{The list of 75 nuclei, whose density distributions are adopted in the PCA procedure in the main text}
    \label{suptable:1}
\end{table}

\section*{Scaling procedure of nuclear densities}

To eliminate the scale effects of nuclear densities arising from variations in mass number $A$ and proton number $Z$ across different nuclei, we normalized density distributions of different nuclei before PCA.
This allows the PCA to focus on extracting the distribution features of nuclear densities, instead of the scaling among different nuclei.
To this end, using $^{40}\text{Ca}$ as the standard, we applied scaling factors simultaneously to the radial coordinate and the density amplitude.
Two scaling factors are introduced.
The radial scaling factor $S_R$ adjusts the proton root-mean-square radius of each nucleus to the reference value of $R_{p}({}^{40}{\rm Ca}) = r_{0} A^{1/3}$ (with $r_{0} = 1.2$ fm and $A=40$), thereby normalizing the spatial extent. 
Simultaneously, the amplitude scaling factor $S_Z$ scales the density integral to renormalize the total proton number to a fixed value of $Z=20$, thus normalizing the amplitude. 
The specific transformation formulas are as follows:
\begin{align}
    R_{p}&=\sqrt{\frac{\int^{\infty}_{0}r^{2}\rho_{p}(S_{R}\cdot r)4\pi r^{2}{\rm d}r}{\int^{\infty}_{0}\rho_{p}(S_{R}\cdot r)4\pi r^{2}{\rm d}r}} =r_{0}A^{1/3}, \label{supeq:1} \\
    Z&=\int^{\infty}_{0}S_{Z}\cdot \rho_{p}(r)4\pi r^{2}{\rm d}r=20 .\label{supeq:2}
\end{align}
This procedure ensures that the established set of density basis functions is efficient for all atomic nuclei.
It should be noted that the density distributions represented by the obtained PCA bases remain in the normalized scale.
One can easily apply the inverse transformation of Eqs.~\eqref{supeq:1} and \eqref{supeq:2} using the scaling factors $S_{R}$ and $S_{Z}$ to obtain the density distributions in the physical scale.

\section*{Performance of PCA method versus the number of PCs}
\begin{figure}[!h]
  \centering
  \includegraphics[width=0.90\linewidth]{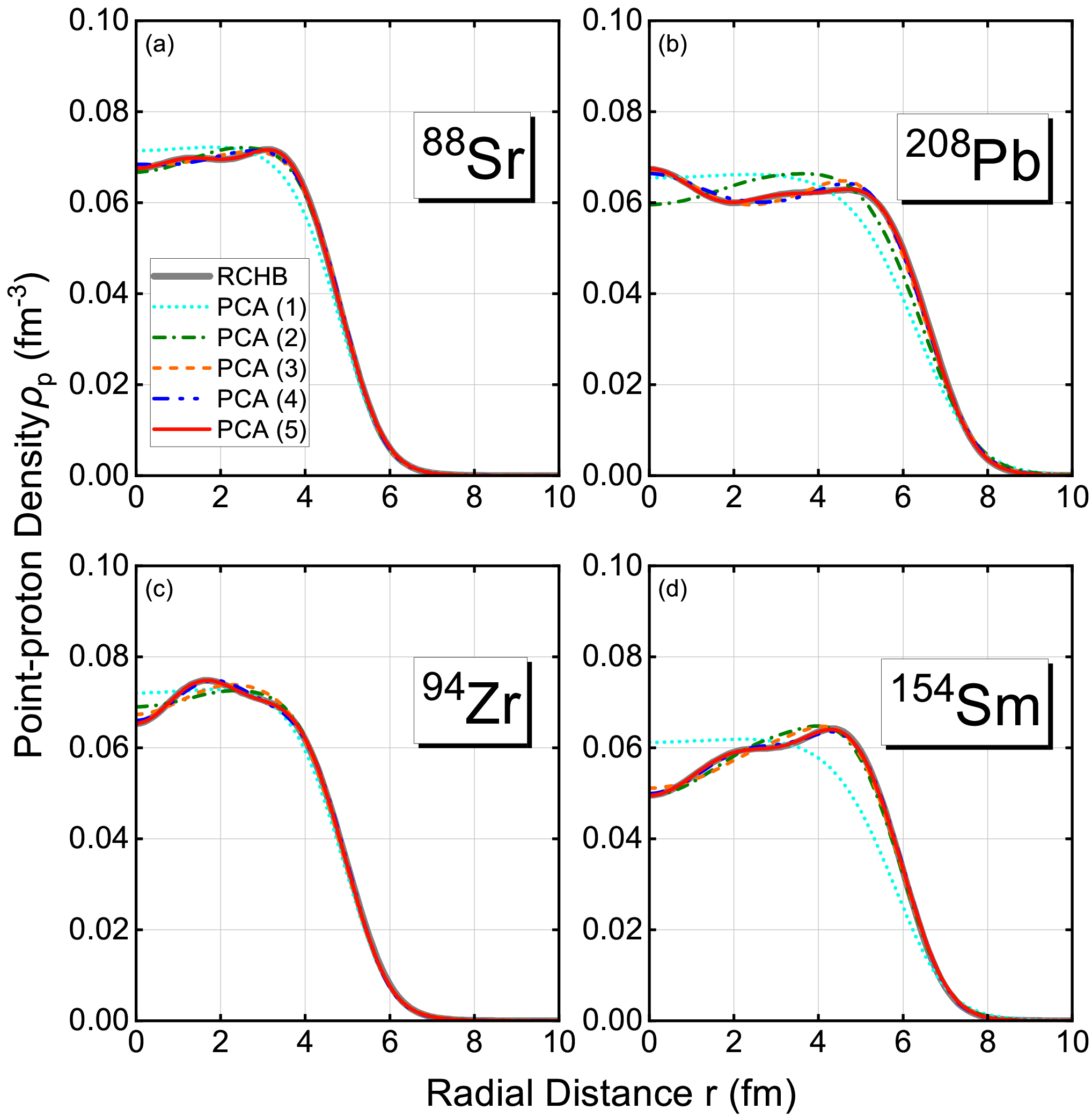}
  \caption{Comparison of the PCA method for representing the proton density distributions of $^{88}$Sr, $^{208}$Pb, $^{94}$Zr, and $^{154}$Sm from RCHB calculations with varying numbers of principal components.}
  \label{Fig.12}
\end{figure}

As can be seen in Fig.~\ref{Fig.12}, the inclusion of more principal components in the PCA method leads to a more accurate representation of the densities, and five principal components can already nicely reproduce the RCHB densities.

\section*{Performance of PCA, FB, and SOG methods under logarithmic coordinates}
\begin{figure}[!h]
  \centering
  \includegraphics[width=0.90\linewidth]{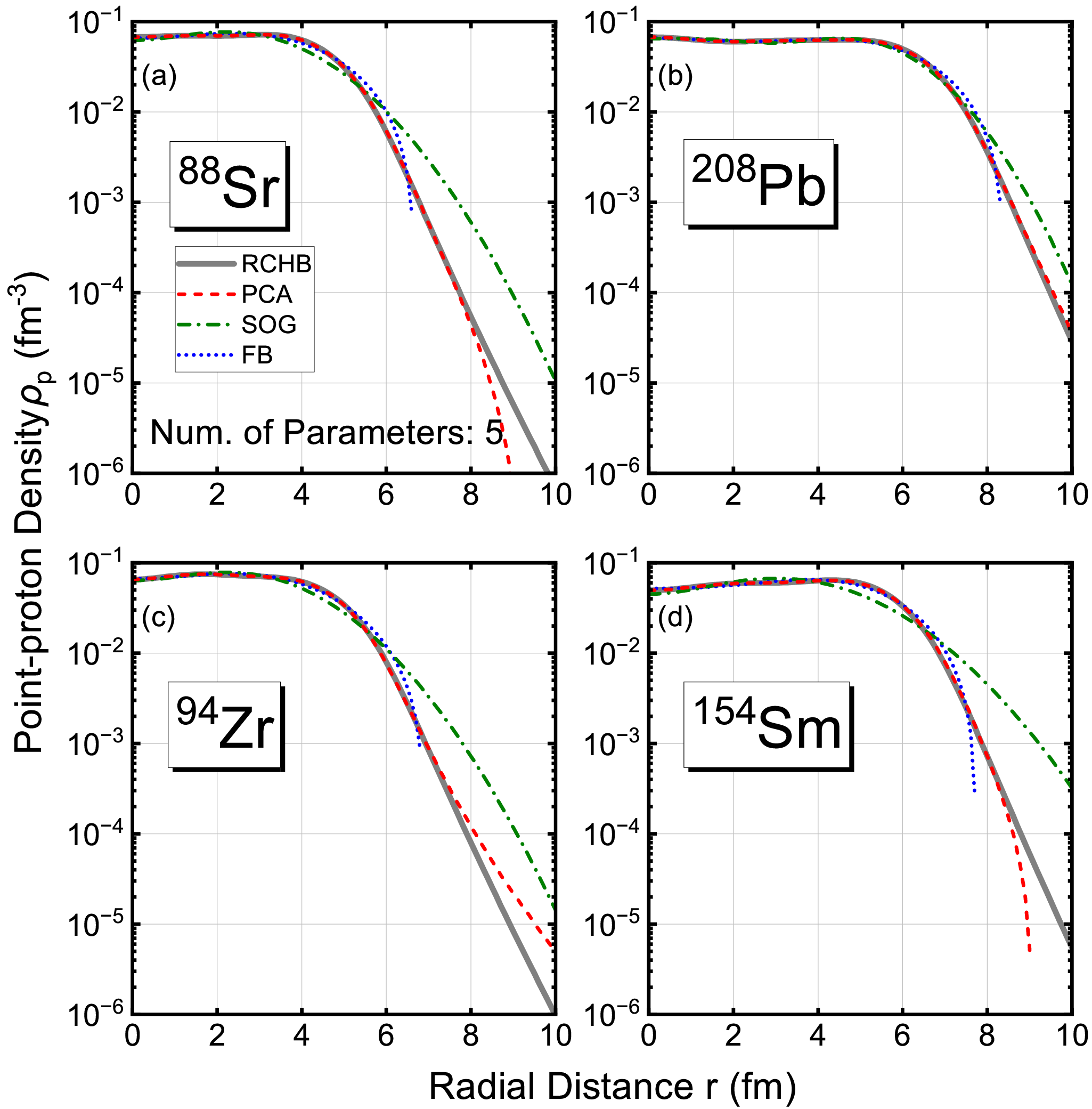}
  \caption{The same as in Fig.~3 of the main text, but presented in logarithmic coordinates. Note that minor negative density values may arise in the basis expansion methods, i.e., PCA and FB. The negative values are not shown in the plot.
  }
  \label{Fig.10}
\end{figure}

\begin{figure}[!t]
  \centering
  \includegraphics[width=0.90\linewidth]{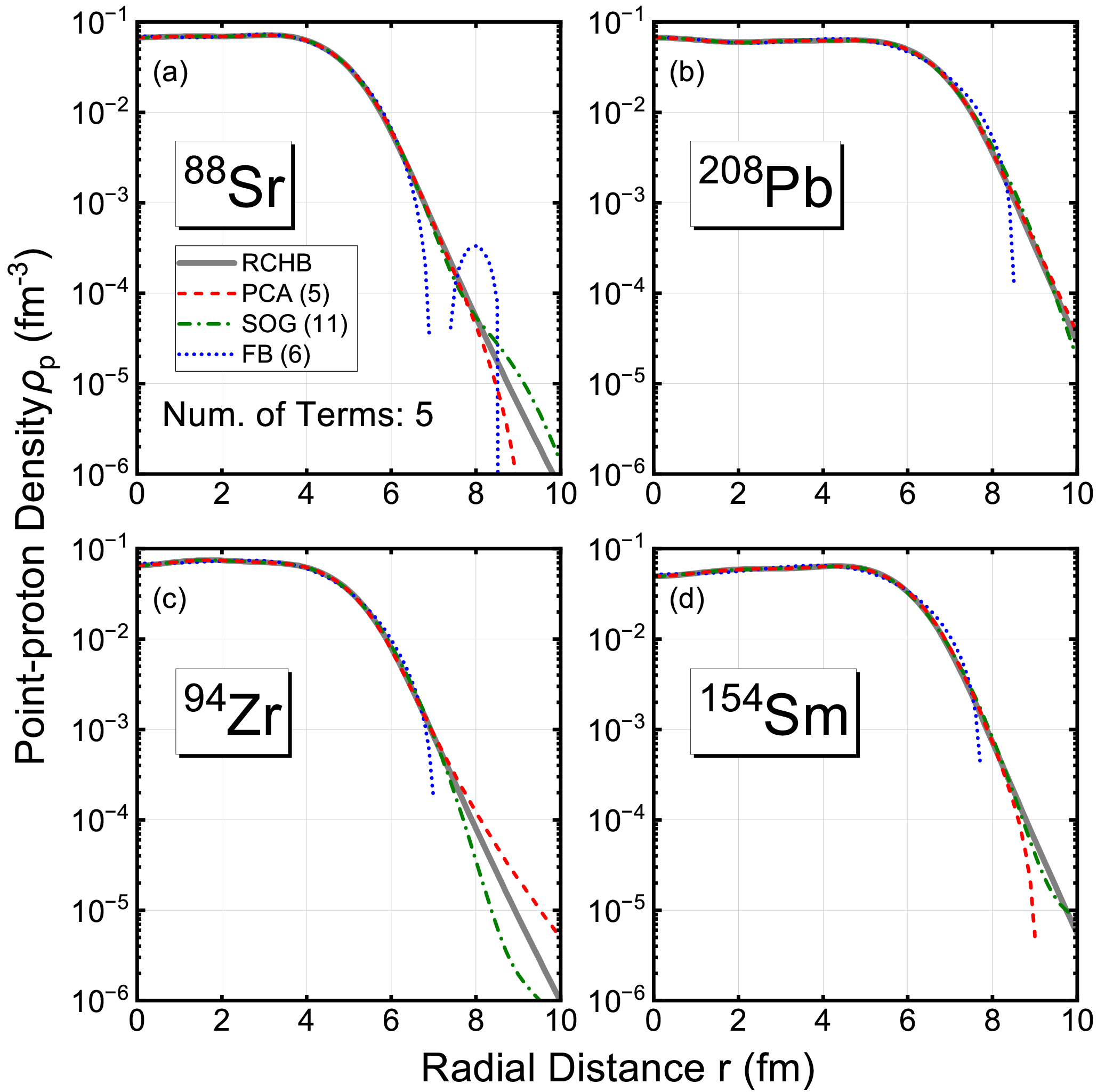}
  \caption{The same as in Fig.~4 of the main text, but presented in logarithmic coordinates. Note that minor negative density values may arise in the basis expansion methods, i.e., PCA and FB. The negative values are not shown in the plot. }
  \label{Fig.11}
\end{figure}

The same densities as in Figs.~3 and 4 of the main text are represented in log scale in Figs.~\ref{Fig.10} and ~\ref{Fig.11}.
This facilitates a detailed comparison of the density tails.
Note that logarithmic coordinates cannot display negative values. 
Since basis expansion methods such as FB and PCA do not guarantee the positivity, curve discontinuities appear in some subfigures.
As can be seen in Fig.~\ref{Fig.10}, the PCA method represent the density tails better than the SOG and FB methods.
The SOG method can provide a reasonable representation of the density tails when the number of fitting terms is 5, as shown in Fig.~\ref{Fig.11}.
Note that the SOG method utilizes a significantly larger number of actual fitting parameters than the PCA and FB methods when the number of fitting terms is equal.
As can be see in both Figs.~\ref{Fig.10} and \ref{Fig.11}, the FB method does not work well to represent the density tails.
Negative density values appear in the region where $r>7$ fm, and the method gives rise to unphysical oscillatory behaviors attributed to the use of high-order basis functions.

\section*{Universality of the PCA density bases}
\begin{figure}[!h]
  \centering
  \includegraphics[width=0.90\linewidth]{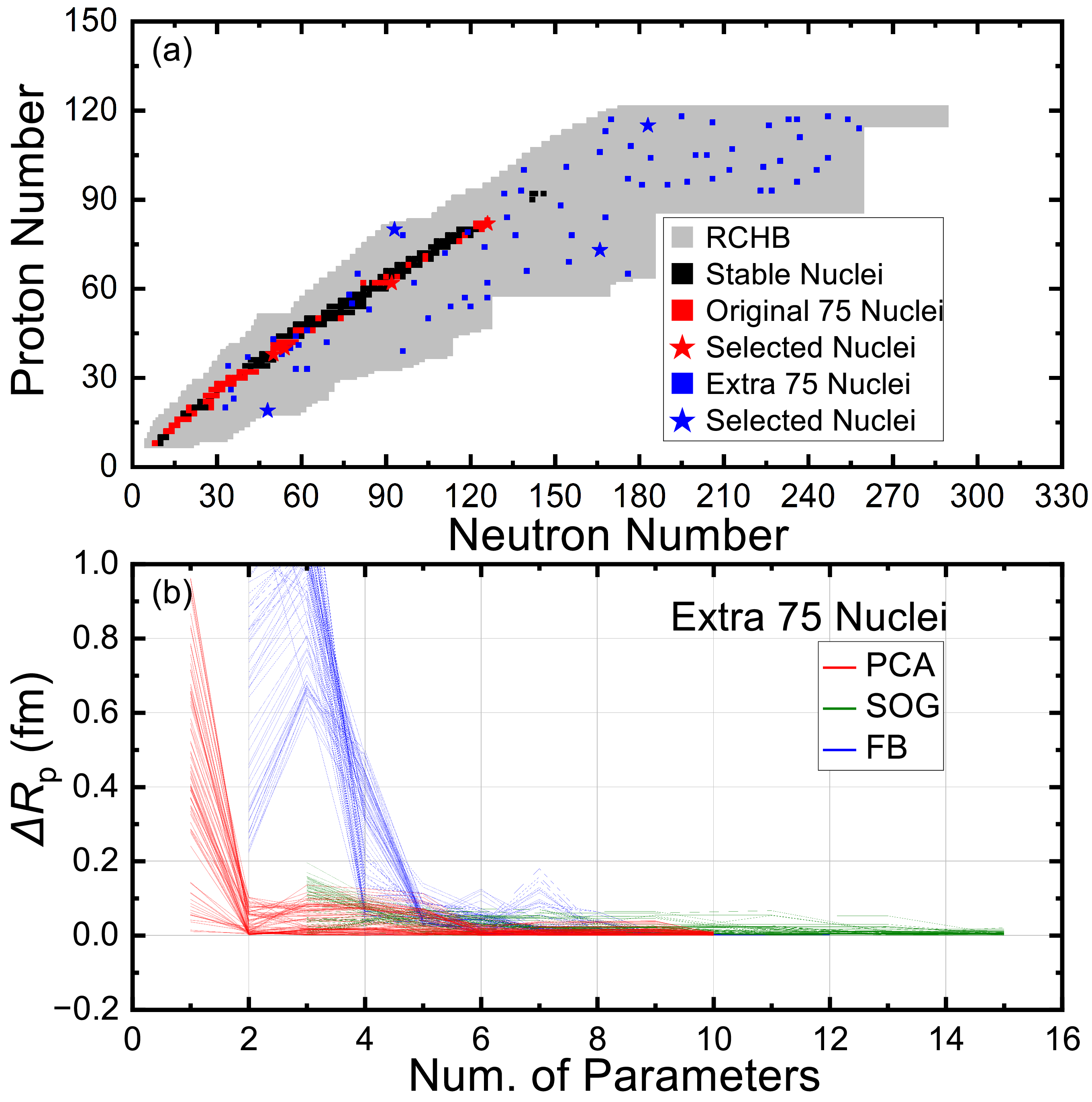}
  \caption{(a) The nuclear chart shows related nuclei adopted in validating the performances of PCA, FB, and SOG methods. 
  Black squares denote stable nuclei. 
  Red squares represent the 75 nuclei used to extract the PCA bases. 
  Red pentagrams denote four randomly selected nuclei: $^{88}$Sr, $^{208}$Pb, $^{94}$Zr, and $^{154}$Sm, which are used in Fig.~3 of the main text. 
  Blue squares denote extra randomly pick-up 75 nuclei outside the dataset of original nuclei that been used to extract the PCA bases.
  Blue pentagrams denote four nuclei randomly chosen from these extra nuclei: $^{67}$K, $^{173}$Hg, $^{239}$Ta, and $^{298}$Mc.
  The gray background represents the bound nuclei predicted by the RCHB theory.
  (b) The same as in Fig.~5, but for 75 extra nuclei.
  }
  \label{Fig.7}
\end{figure}
\begin{figure}[!t]
  \centering
  \includegraphics[width=0.95\linewidth]{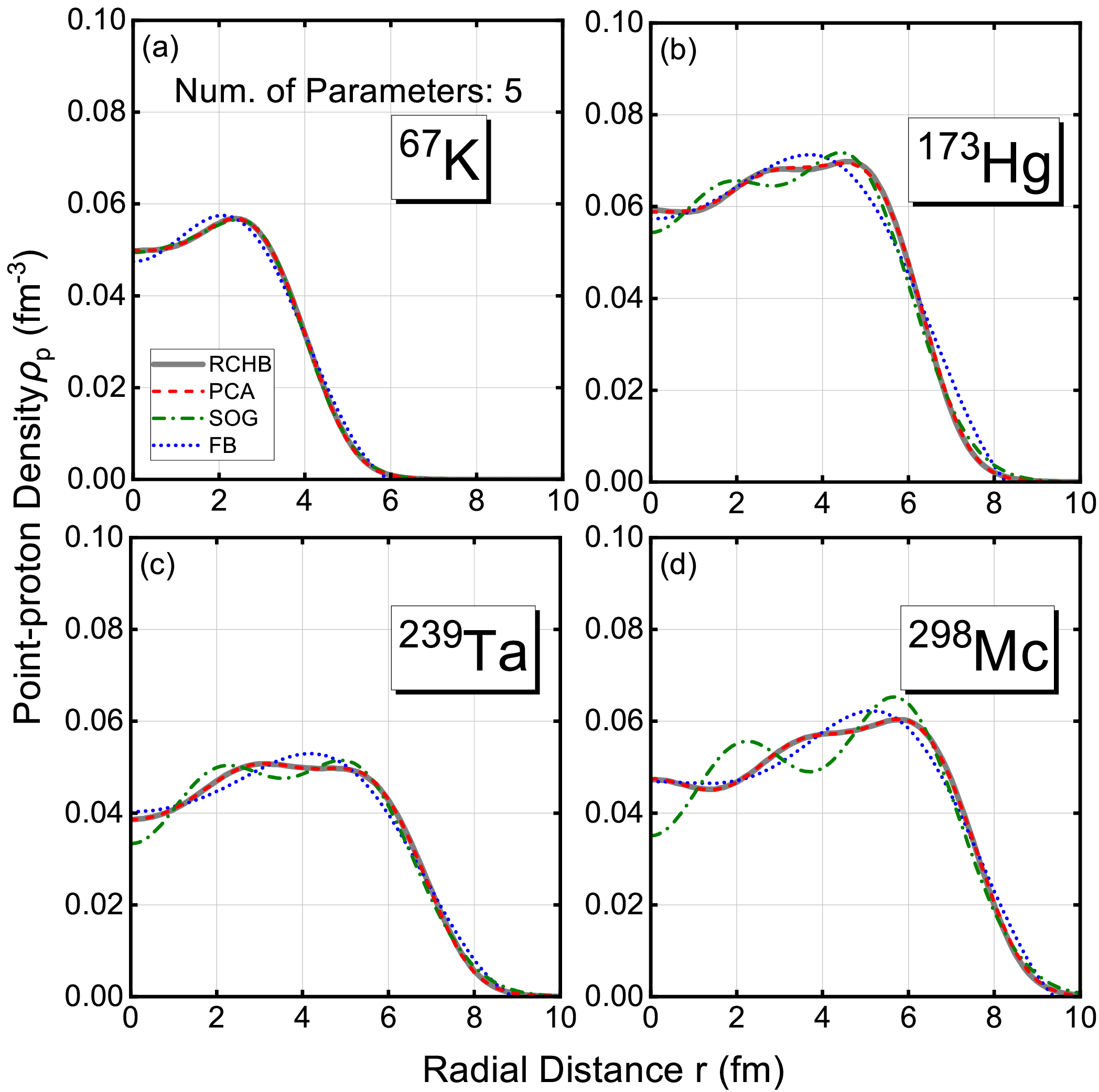}
  \caption{The same as in Fig.~3 of the main text, but for $^{67}$K, $^{173}$Hg, $^{239}$Ta, and $^{298}$Mc.
  }
  \label{Fig.8}
\end{figure}
\begin{figure}[!t]
  \centering
  \includegraphics[width=0.95\linewidth]{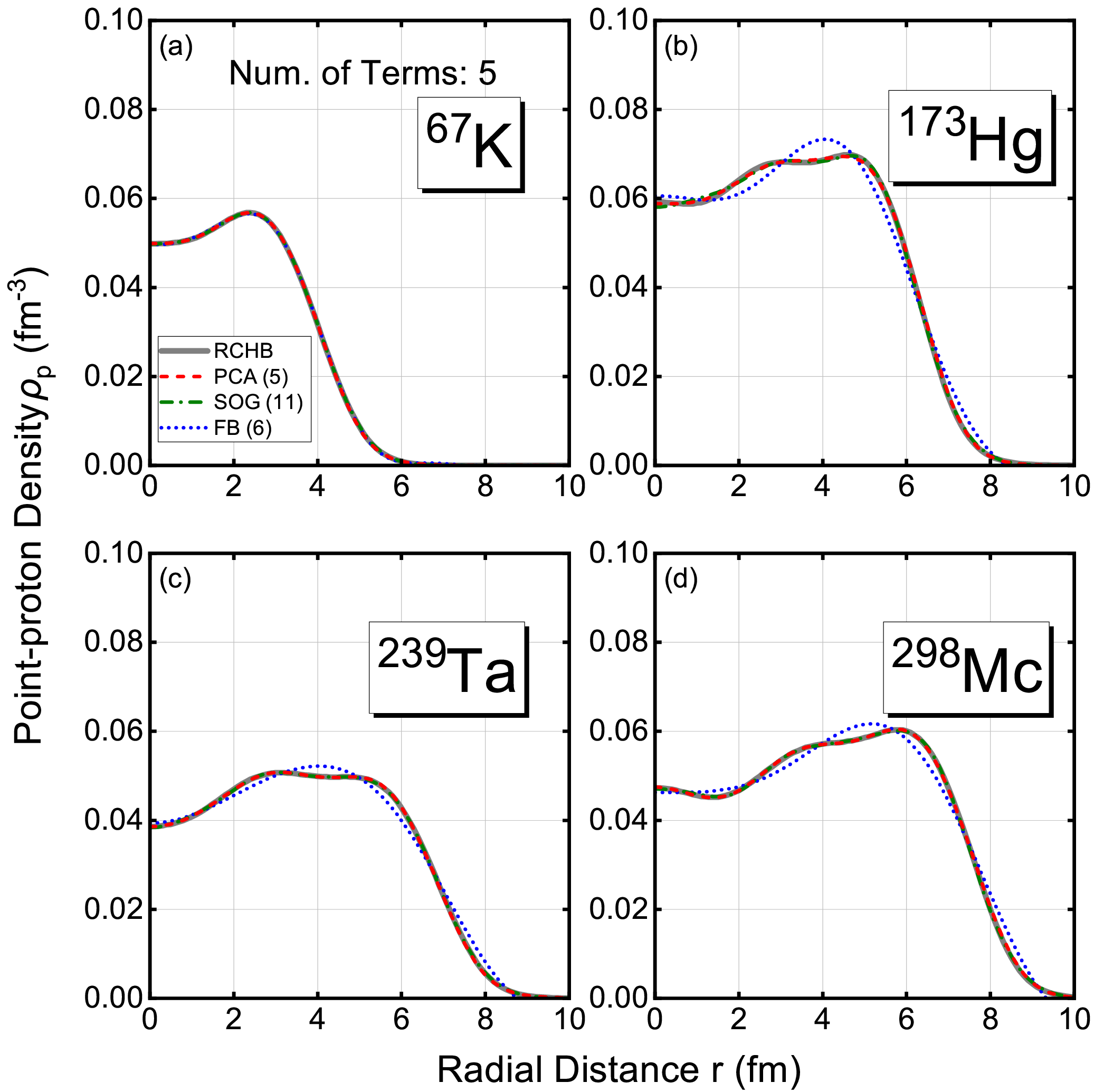}
  \caption{The same as in Fig.~4 of the main text, but for $^{67}$K, $^{173}$Hg, $^{239}$Ta, and $^{298}$Mc.
  }
  \label{Fig.9}
\end{figure}

75 extra nuclei, outside the dataset of original ones used in the PCA procedure, are randomly picked up to 
validate the universality of the PCA bases in representing nuclear densities.
These nuclei are shown in Fig.~\ref{Fig.7}~(a) together with the original 75 nuclei.
One can see from Figs.~\ref{Fig.7}~(b), \ref{Fig.8}, and \ref{Fig.9} that, the PCA method works well and better than both the FB and SOG methods, even for nuclei outside the original 75 nuclei used to extract the PCA bases.
This indicates the universality of the obtained PCA bases in representing nuclear densities across the nuclear chart.

